\documentclass[prb,twocolumn,superscriptaddress,showpacs,amssymb]{revtex4-1}

\usepackage{soul}

\usepackage{ifthen}
\newboolean{standaloneFlag}
\setboolean{standaloneFlag}{true}

\newcommand\standaloneBib{
  \ifthenelse{\boolean{standaloneFlag}}
             {\printbibliography}{}}

\makeatletter
\newcommand\journal{\@journal}
\makeatother

\usepackage{pdftexcmds}

\makeatletter
\ifdefined\@journal

\ifnum\pdf@strcmp{\journal}{prl}=0 %
  \renewcommand{\section}[1]{\@startsection{section}{1}{\z@}%
                             {-3.25ex\@plus -1ex \@minus -.2ex}%
                             {-1.5ex \@plus -.2ex}
                             {\indent\normalfont\normalsize\bfseries\itshape}{#1:--}}
\else
\fi
\fi

\makeatother

\usepackage{dcolumn}{}
\usepackage{bm}
\usepackage{amsfonts,amssymb,amsmath} 

\usepackage[dvipsnames]{xcolor}

\usepackage{tikz}

\usepackage{epstopdf}
\usepackage{mathbbol}

\usepackage{amsthm}

\usepackage[breaklinks=true]{hyperref}
\usepackage{breakcites}

\usepackage{slashed}

\theoremstyle{plain}

\theoremstyle{definition}

\theoremstyle{remark}

\newcommand{\be}{\begin{equation}}
\newcommand{\ee}{\end{equation}}
\newcommand{\bq}{\begin{eqnarray}}
\newcommand{\eq}{\end{eqnarray}}

\newcommand{\rf}[1]{(\ref{#1})}

\newcommand{\1}{1\!\!1}

\newcommand{\tr}{\mathrm{tr}}
\newcommand{\ket}[1]{\left |#1 \right\rangle}

\usepackage{wasysym}
\usepackage{soul}

\usepackage{hyperref}

\usepackage{tikz}

\usepackage{verbatim}

\usetikzlibrary{positioning,shapes.misc,calc,decorations.markings}
\tikzset{->-/.style={decoration={
      markings,
      mark=at position #1 with {\arrow{latex}}},postaction={decorate}}}
\tikzset{-<-/.style={decoration={
      markings,
      mark=at position #1 with {\arrow{latex reversed}}},postaction={decorate}}}
\usetikzlibrary{shapes.geometric}

\usepackage{amsfonts,amssymb,amsmath}

\usepackage{float}
\usepackage[caption = false]{subfig}
\usepackage{graphicx}

\usepackage{hyperref}

\usepackage{amsfonts,amssymb,amsmath} 
\usepackage{slashed}
\usepackage[dvipsnames]{xcolor}

\usepackage{textcomp}

\begin{document}


\title{Machine Learning Entanglement Freedom Or:\\
How I Learned to Stop Worrying and Love Linear Regression}


\author{Samuel Spillard}
\affiliation{School of Physics and Astronomy, University of Leeds, Leeds LS2 9JT, United Kingdom}

\author{Christopher J. Turner}
\affiliation{School of Physics and Astronomy, University of Leeds, Leeds LS2 9JT, United Kingdom}

\author{Konstantinos Meichanetzidis}
\affiliation{School of Physics and Astronomy, University of Leeds, Leeds LS2 9JT, United Kingdom}


\date{\today}

\begin{abstract}
Quantum many-body systems realise many different phases of matter characterised by their exotic emergent phenomena.
While some simple versions of these properties can occur in systems of free fermions,
their occurrence generally implies that the physics is dictated by an interacting Hamiltonian.
The interaction distance has been successfully used to quantify the effect of interactions in a variety of states of matter via the entanglement spectrum [Nat. Commun. {\bf 8}, 14926 (2017), arXiv:1705.09983].
The computation of the interaction distance reduces to a global optimisation problem whose goal is to search for the free-fermion entanglement spectrum closest to the given entanglement spectrum.
In this work, we employ techniques from machine learning in order to perform this same task.
In a supervised learning setting, we use labelled data obtained by computing the interaction distance and predict its value via linear regression.
Moving to a semi-supervised setting, we train an auto-encoder to estimate an alternative measure to the interaction distance, and we show that it behaves in a similar manner.
\end{abstract}

\maketitle

\section{Introduction}

The interaction distance is a diagnostic measure
of a pure state's non-Gaussianity as it manifests in
its entanglement structure.
In essence, it performs pattern recognition
on entanglement spectra, where the pattern is dictated by the
Fermi-Dirac statistics obeyed by free fermions.
When applied to
ground states of quantum many-body systems,
the results can be counter-intuitive and surprising,
even for well known and extensively studied systems,
such as the quantum Ising chain~\cite{Schultz},
parafermion chains~\cite{FradkinKadanoff,Fendley:2012hw}, and string-net models~\cite{WenBook}.

Computing $D_\mathcal{F}$ involves a non-convex optimisation.
Even if it is \emph{in principle efficient}~\cite{Turner2017},
its computation runtime still remains impractical for entanglement spectra which are
accessible efficiently with current numerical methods~\cite{Verstraete}.
Then a natural question arises;
can we train a model to \emph{predict} the interaction distance
so that one does not need to perform a global optimisation for every input entanglement spectrum.
In principle, the answer is affirmative. However,
the caveat is that the complexity of the training set is too high if we want it to work accurately for \emph{any} spectrum.
With this motivation, we approach case studies found in previous works
from a machine learning perspective.
As we focus on particular models,
the learning model and the training set can be chosen to be simple,
at the cost of generality.
We find that in these special cases,
the machine learning methods reproduce results for the interaction distance
with a noticeable performance improvement.

\section{The Problem of Computing the Interaction Distance}\label{IntDist}

\subsection{Entanglement Spectrum}

The entanglement spectrum is defined as the spectrum
of a mixed state $\rho$ obtained after
biparitioning the domain of a pure state $\ket{\psi}$
into regions $R,\bar R$ and
performing a partial trace over $\bar R$.
We denote the eigenvalues of $\rho$ by $P_k \in [0,1]$,
as it corresponds to a probability distribution $P$.
We also recall the definition of the entanglement energies~\cite{Haldane08}
as ${E_k} = -\log{P_k} \in [0,\infty)$.
An equivalent definition of the entanglement spectrum invokes the Schmidt decomposition
of a pure state $\psi = \sum_k \xi_k \ket{\psi}^R_k \ket{\psi}^{\bar R}_k$
onto independent orthonormal bases supported on each complementary region.
Then we have that $P_k = \xi_k^2$.

\subsection{Interaction Distance}

To quantify the non-Gaussianity of a mixed state $\rho$,
we define the interaction distance~\cite{Turner2017}
as $D_\mathcal{F}(\rho) = \min_{\sigma\in\mathcal{F}} D(\rho,\sigma)$.
This is the minimal trace distance, $D(\rho,\sigma)$, between $\rho$
and the manifold ${\cal F}$, which contains
all free-fermion reduced density matrices, $\sigma$.
In other words, the interaction distance quantifies
the non-applicability of Wick's theorem for the state $\rho$.

It was proven~\cite{Turner2017} that
since relative rotations between matrices can only increase the trace distance,
the interaction distance
can be expressed exclusively in terms of the spectra of those matrices, i.e. the entanglement spectra.
In other words, the trace distance reduces to the $1$-norm $D(X,Y)=\sum_i \frac{1}{2}\left| X_i-Y_i \right|$, with $0 \leq D(X,Y)\leq 1$, for probability distributions $X,Y$.

For the purposes of this work,
it is useful to formulate
the interaction distance in two equivalent ways, distinguished by the space in which the minimisation takes place, i.e. probability- versus energy-space,
\bq
  \label{eq:DF}
    D_\mathcal{F}(P)&=&\frac{1}{2}\min_{s} \sum_k \left| P_k - P^\mathrm{f}_k (s) \right| ,\\\nonumber
    D_\mathcal{F}(E)&=&\frac{1}{2}\min_{\epsilon} \sum_k \left| e^{-E_k} - e^{-E_k^\mathrm{f}(\epsilon)} \right| ,
\eq
where
$\mathrm{f}$ indicates a spectrum with a free fermion structure generated by the
\emph{polynomially large} single-body sets $s$ and $\epsilon$, to be defined in
Eq.\rf{eq:freespec}, and the spectra are rank ordered, $E_k\leq E_k+1$ and $E^\mathrm{f}_k\leq E^\mathrm{f}_k+1$ or $P_k\geq P_k+1$ and $P^\mathrm{f}_k\geq P^\mathrm{f}_k+1$.

Intuitively, $D_\mathcal{F}$ is dominated by the low-lying part of the entanglement Energy spectum and it reveals the correlations between the effective quasiparticles emerging from interactions~\cite{Haldane08}.
Hence, $D_\mathcal{F}$ is expected to be stable under perturbations that do not cause phase transitions~\cite{Turner2017}.

\subsection{Interaction Distance as an Inverse Problem}
\label{sec:expand}

In order to study $D_\mathcal{F}$ with machine learning methods,
we formulate its estimation as an inverse problem.
A free fermion spectrum is defined to be one that is created by the expand map $\mathcal{E}$, which takes as input a set of single-body probabilities or energies  and outputs a probability or energy spectrum which obeys the free fermion structure,~\cite{Haldane08,PeschelEisler}
\bq
\mathcal{E}_\mathrm{P} &:& s\in \mathbb{R}^N_{<} \to P^\mathrm{f} (s)\in \mathbb{R}^{2^N}_{>}, \\\nonumber
\mathcal{E}_\mathrm{E}  &:& \epsilon\in \mathbb{R}^N_{<} \to E^\mathrm{f}(\epsilon) \in  \mathbb{R}^{2^N}_{<} .
\eq
The free-fermion spectra in probability space and entanglement energy space
have the forms
\bq \label{eq:freespec}
P^\mathrm{f}(s)&=&\mathrm{sort}_{\mathrm{desc}}~ \otimes_{i=1}^N \left(\frac{1}{2}+s_i,\,\frac{1}{2}-s_i\right) \\\nonumber
E^\mathrm{f}(\epsilon)&=&\mathrm{sort}_{\mathrm{asc}} ~E_0+\oplus_{i=1}^N \{0,\epsilon_i\}
\eq
with $0\leq s_i\leq\frac{1}{2}$,
and $E_0=-\sum_i \log Z_i$ with $Z_i=1+e^{-\epsilon_i}$.
Equivalently we can write
$P^\mathrm{f}=\mathrm{sort}_{\mathrm{desc}}~\otimes_{i=1}^N \frac{1}{Z_i}(1,p_i)$
with $p_i=e^{-\epsilon_i}$.
However, the  $s$-parametrisation of free-fermion probability spectra
in Eq.\rf{eq:freespec}
produces inherently normalised $P^\mathrm{f}$ spectra and
makes convenient the computation of $D_\mathcal{F}(P)$.

The solution to the problem of computing $D_\mathcal{F}$ amounts to finding 
the weak inverse of the expand map, which minimises the trance distance for input outside the image of expand.
We denote this generalised inverse as $\mathcal{E}^g$.
If such an inverse exists it has the properties
\bq
\mathcal{E}^g \circ \mathcal{E} \circ  \mathcal{E}^g &=& \mathcal{E}^g ,\\\nonumber
\mathcal{E} \circ  \mathcal{E}^g \circ  \mathcal{E} &= &\mathcal{E}.
\eq
From this perspective, we can write Eq.\rf{eq:DF} as
\bq
  \label{eq:DF_expand}
  D_\mathcal{F} (P) &=& \frac{1}{2}\min_{\mathcal{E}^g_P } \left|\left|  \mathcal{E}_P ( \mathcal{E}_P^g(P))  - P \right|\right|_1  \\\nonumber
  D_\mathcal{F} (E) &=& \frac{1}{2}\min_{\mathcal{E}^g_E } \left|\left| e^{- \mathcal{E}_P ( \mathcal{E}_E^g(E)) } - e^{- E} \right|\right|_1  .
\eq

\subsection{AutoEncoder Perspective} \label{sec:AEPerspective}

Ideally, one would
train a deep neural network,
for example an AutoEncoder (AE)~\cite{Hinton504} that learns the entanglement freedom structure
from a dataset of free states.
The map $\mathcal{E}$ is in general non-linear in the single-body input set.
In an autoencoder setup, we have the
Coding process $\mathcal{C}$, which maps an input $X$
to the latent layer, and the Decoding $\mathcal{D}$ process which maps from the latent layer to the output,
\be\label{eq:autoencoder}
X \rightarrow \mathcal{C}(X) \rightarrow \mathcal{D}(\mathcal{C}(X)). 
\ee
The goal of the training step for the AE is then to minimize $D(\mathcal{D}(\mathcal{C}(X)),X)$ for all $X$,
where $X$ are free states, by updating $\mathcal{C}$ and $\mathcal{D}$ appropriately.
Thus, $\mathcal{C}$ would learn to implement $\mathcal{E}^g$
and $\mathcal{D}$ would learn $\mathcal{E}$.
Then for a new input spectrum $Y$,
the surrogate for the interaction distance is defined as
\be \label{eq:DF_AE}
D^\mathrm{AE}_\mathcal{F}=D(\mathcal{D}(\mathcal{C}(Y)),Y).
\ee
Intuitively, the interaction distance in this case measures how wrong
an autoencoder is when it recognises freedom in a spectrum it has not encountered during training.
However,
obtaining a representative training set of all free states $X\in\mathcal{F}$ is a hard task.
In the next section we do some elementary data analysis on the classes of spectra we consider and in the following sections we employ simple regression methods that estimate the interaction distance in specific contexts.

\section{PCA of spectra}

\begin{figure}[t]
\centering
\includegraphics[width =\columnwidth]{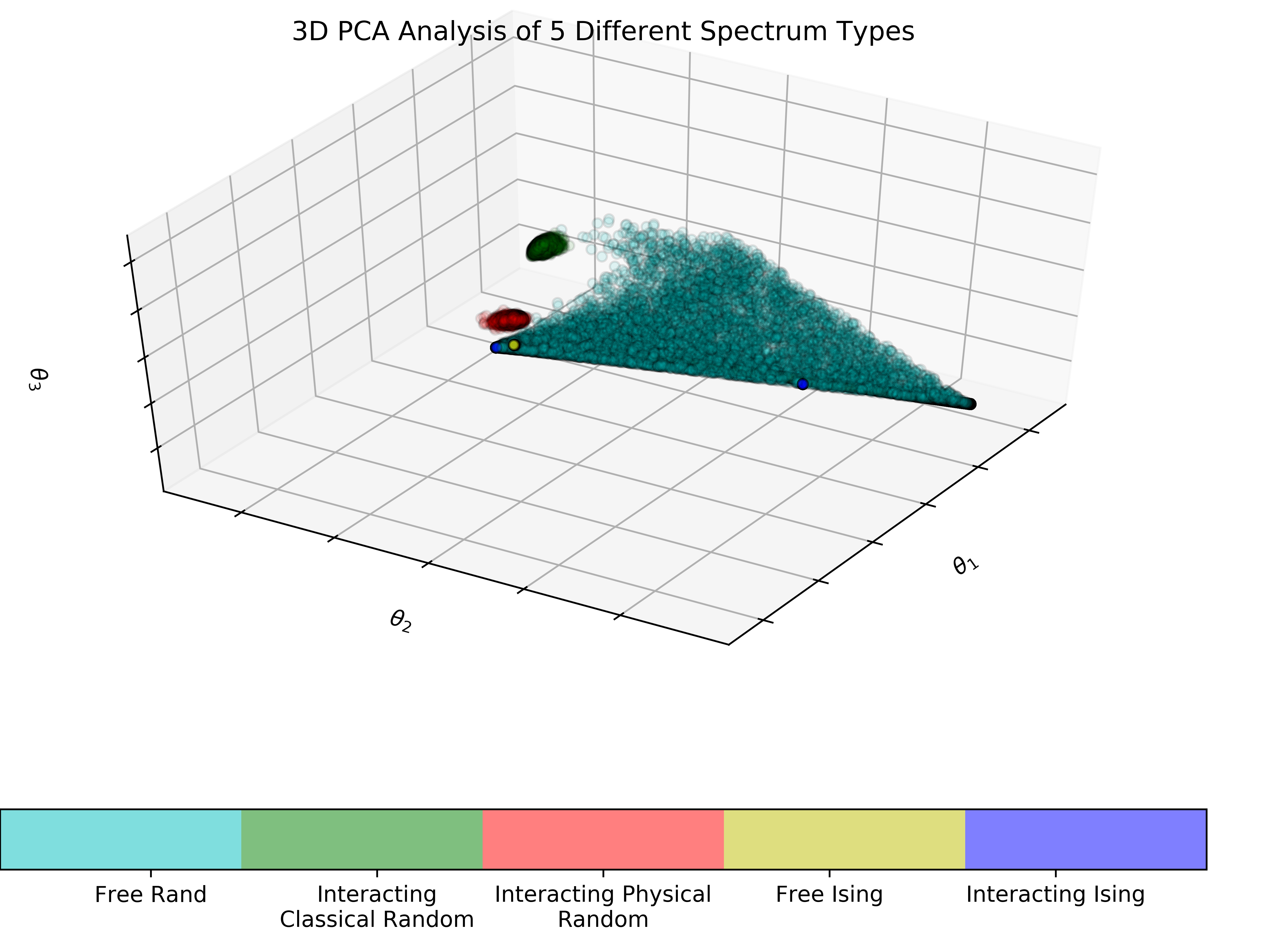}
\caption{PCA of entanglement spectra of different origin (see  Appendix \ref{app:spectracreation}) and thus in general different entanglement patterns.
The data is projected on the three principal vectors $\theta_1, \theta_2, \theta_3$ of highest weight.
Interacting spectra are separated from free or effectively free spectra.
}
\label{fig:PCAspectra}
\end{figure}

Here we present a data analysis of entanglement spectra we consider throughout this work.
Collecting all such spectra of a certain size into a matrix and performing PCA,
we represent this dataset in the space spanned by the first three principal vectors.
In Fig.\ref{fig:PCAspectra} we see how spectra with different properties are clustered,
and we observe that it is reasonable to expect that a freedom classifier can be defined.

\section{Linear Approaches for Specific Case Studies}

We now focus our study to specific sets of entanglement spectra
arising from condensed matter systems of interest to the community,
such as the the paradigmatic quantum Ising chain and Abelian topological models.
To make progress, we consider reducing our learning model's complexity and thus its generality for the sake of lowering the complexity of the training set needed.
To this end, we make a linear approximation in the inverse problem of computing the interaction distance by viewing the expand map as a linear transformation,
\be \label{eq:linearexpand}
\mathcal{E}_X  \sim \slashed{\mathcal{E}}_X ,
\ee
whose form is specified by the problem at hand.

\subsection{Random States} \label{sec:randomstates}

We begin by considering random entanglement spectra,
which are generated as described in Appendix \ref{app:spectracreation}.
The linear approximation to the expand map is done by ignoring the distinction between sets and ordered sets (multisets) in the entanglement energy space,
\be \label{eq:linearexpand}
\slashed{\mathcal{E}}_{E} :  \epsilon\in \mathbb{R}^N \rightarrow E^\mathrm{f}(\epsilon)\in \mathbb{R}^{2^N}.
\ee
We purposefully choose to work in $E$-space, as the linear approximation is not viable in $P$-space.
In the parameterisation of $\mathcal{F}$ by the single-particle energies,
we may take linear combinations of them and up to reorderings
get another single-particle spectrum.
For the many-body spectrum, which is the expansion of some single-body energy levels,
the expansion assigns an occupation pattern to each many-body level.
If you restrict to the subset of single-particle energies that expand to spectra with the same ordering of these occupation patterns with respect to the energy ordering,
then expand is a linear map and accordingly has a linear weak inverse.
Geometrically, in the energy parameterisation the manifold $\mathcal{F}$ is piecewise linear, whereas
in terms of the energy variables, $\mathcal{F}$ is locally a hyperplane.
This is our main justification for using the energy-space parametrisation for the linear regression problem.
For the probabilities, on the other hand, the manifold $\mathcal{F}$ is curved.
Thus, attempts to contain it within a hyperplane, or to find a hyperplane contained within it are likely to form gross approximations.

The least-squares linear regression in energy-space has a number of flaws.
Firstly, as discussed the model can't capture the ordering structure.
Secondly, the least squares cost function weights all deviations in energy equally.
Howerver, for the trace distance cost function, large variations in high entanglement energy levels, which are highly penalised by the least-squares solution, would be an insignificant variation in probability.
Similarly, a small energy variation in the low lying energy levels would be more important.

Since now $\slashed{\mathcal{E}}_{E}$ is a linear transformation, it has a $ 2^N \times N $ matrix representation with columns carrying bit-strings corresponding to occupation patterns labelling the Fock basis states on the independent fermionic modes.
This matrix acts on a column vector containing the single-body energies $\epsilon_i$, $i=1,\dots,N$
and results in a column vector containing the many-body energies $E_k$, $k=1,\dots,2^N$.

The map $\slashed{\mathcal{E}}_E$ is linear
and has full column-rank,
it therefore has a Moore-Penrose pseudoinverse
with the property $\slashed{\mathcal{E}}_{E}^g\slashed{\mathcal{E}}_{E}=\1$.
We then apply linear regression to infer a design matrix $F$,
which is identified with the linear map $\slashed{\mathcal{E}}_{E}^g$,
such that $\epsilon = F E + \delta$. We use the
least squares method so that $\left|\left|\epsilon - F E \right|\right|_2 =\delta^2$
is minimised.
Here the set $\epsilon$ contains the single-body energies that are computed by our
algorithm~\cite{Turner2017} for $D_\mathcal{F}$
for the entanglement spectra $E$ from random states,
$\epsilon=\mathrm{argmin} D(E,E^\mathrm{f}(\epsilon))$.
The algorithm first builds an initial guess for the free spectrum by examining the input spectrum.
Then a local optimisation is performed, for example via the Nelder-Mead algorithm. Finally,
a basin-hopping Monte Carlo technique can be introduced to ensure that enough local minima are visited and that the global one is reached with high probability. The number of basins is a parameter left free. Here,
the basin-hopping is turned off and the minimisation is local, with the initial guess
constructed as described in Ref~\onlinecite{Turner2017}.
This approach can thus be called supervised, as $D_\mathcal{F}$ is the label of each $E$.
The matrix $F$ is expected to implement $\mathcal{E}^g$, i.e. the linear approximation to the weak inverse of the expand map.
Then the estimated interaction distance is $D_\mathcal{F}^\mathrm{est}(E) = D( e^{-E},  e^{-E^\mathrm{f}( FE )} )$.

\begin{figure}[t!]
\begin{center}

 \includegraphics[width=\columnwidth]{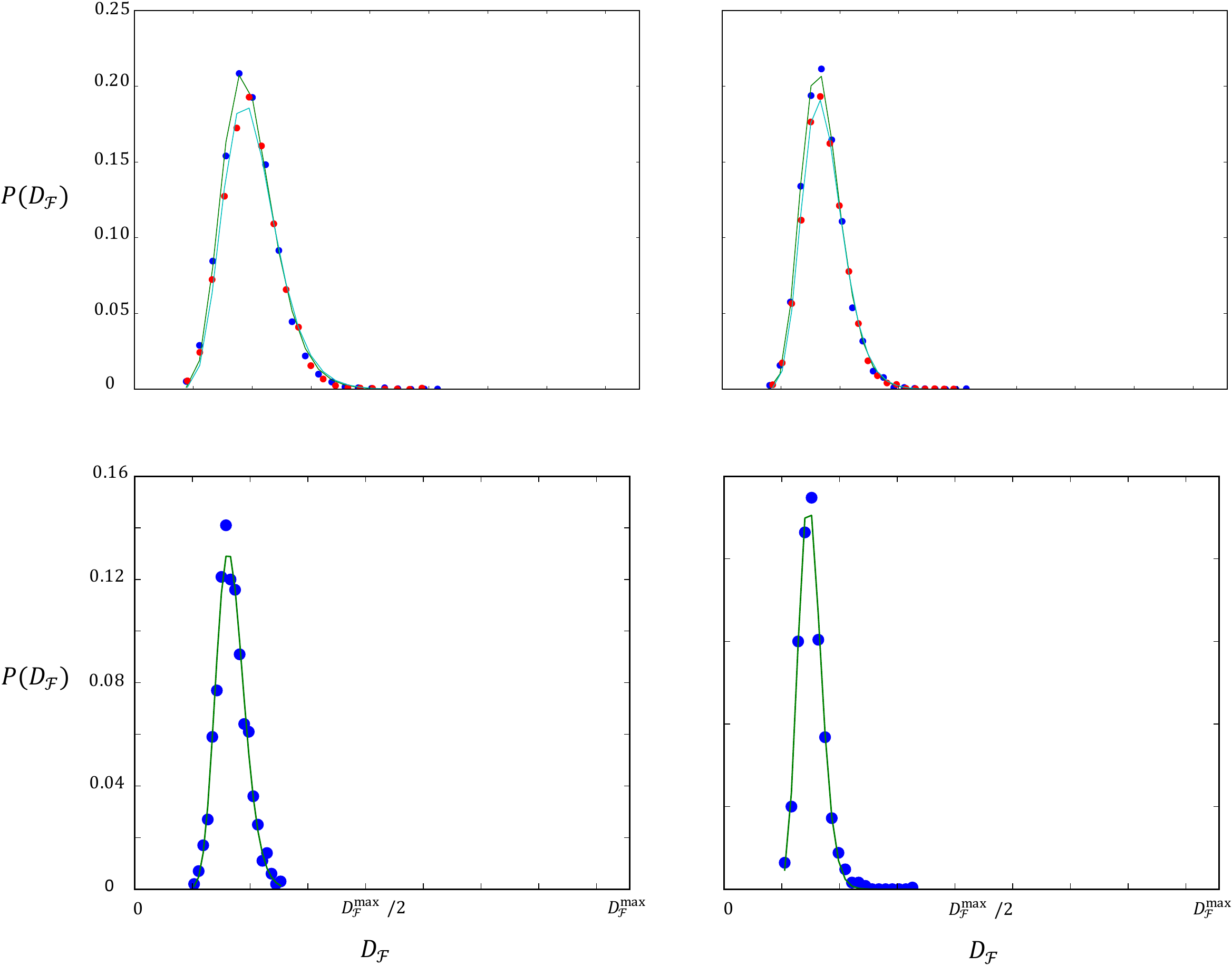}
 \includegraphics[width=.87\columnwidth]{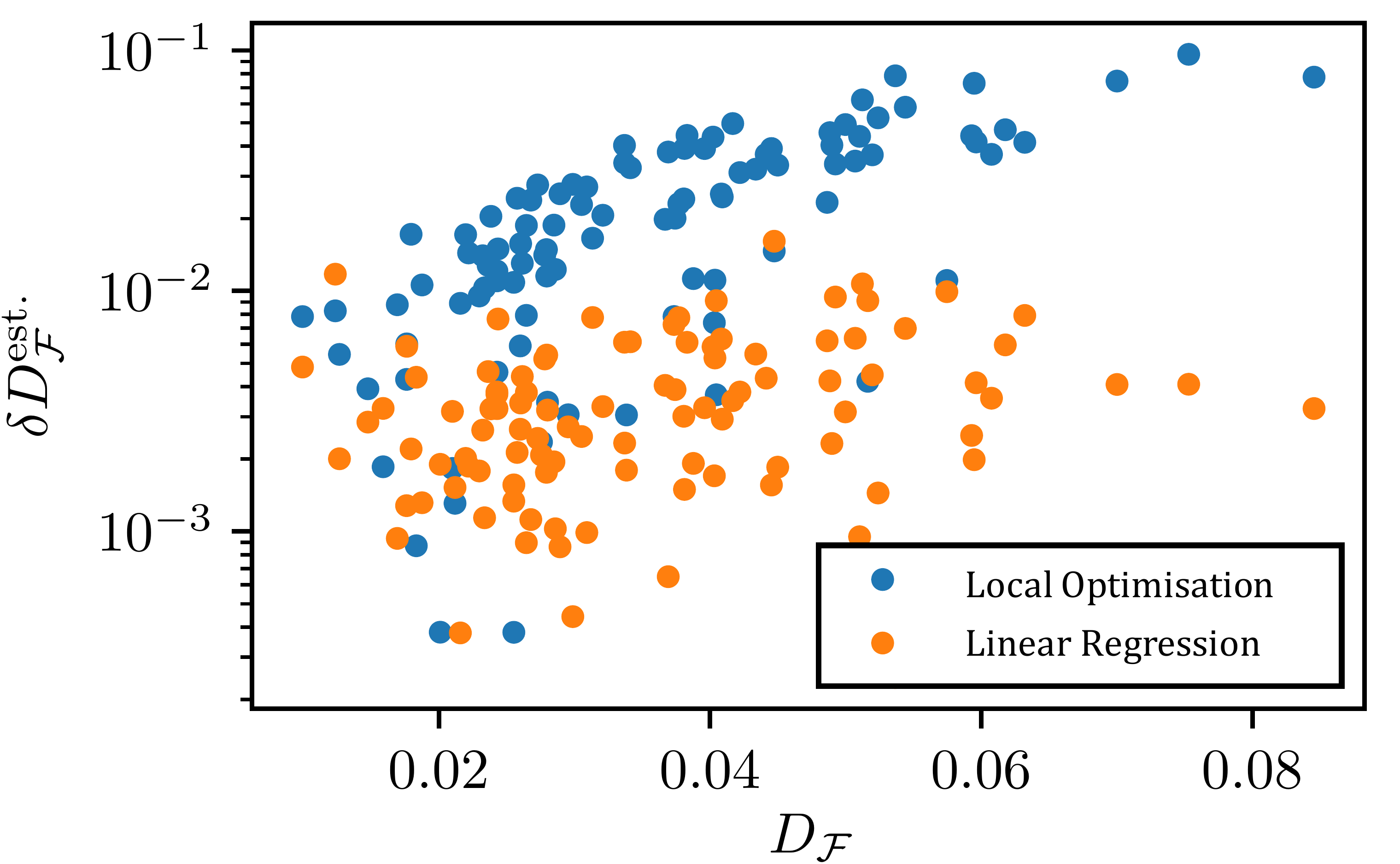}

\end{center}

\caption{
(Top) $P(D_\mathcal{F})$ 
over $5\cdot 10^3$ states for $q={8}$ and $\psi_j$ drawn from the power-law distribution $\delta x^{-\delta}$.
Amplitudes sampled as
$\psi_j\in\mathbb{R}$ (Left) and $\psi_j\in\mathbb{C}$ (Right),
and ${P(D_\mathcal{F})}$ is fitted with ${P_\text{LN}}$,
with dots corresponding to ${\delta=2}$ (blue) and ${\delta=20}$ (red).
(Middle) $P(D_\mathcal{F})$
for $10^3$ states on $q=10$ qubits
with
$\psi_j\in\mathbb{R}$ (Left) and $\psi_j\in\mathbb{C}$ (Right)
drawn from $\mathcal{N}(0,1)$.
Again, $P(D_\mathcal{F})$ is fitted well with ${P_\text{LN}}$.
The most common value for the interaction distance is comparable in all cases.
(Bottom) Linear regression accuracy $\delta D^\mathrm{est}_\mathcal{F}=\left| D^\mathrm{est}_\mathcal{F}-D_\mathcal{F}\right|$ as the difference of its prediction from direct computation of the interaction distance.
}
\label{fig:LR_RandomStates}
\end{figure}

The distribution of $D_\mathcal{F}$ over random states
which is fitted well by a log-normal distribution, $P_\text{LN}(X,a,b)=\frac{1}{ x b \sqrt{2\pi} } e^{- \frac{(\ln{X}-a)^2}{2 b^2}}$,
and the accuracy of this linear method are shown in Fig.\ref{fig:LR_RandomStates}.
We observe that the linear regression estimation is able to perform unexpectedly better than
the local minimisation version of our current algorithm.
Note that direct computation of $D_\mathcal{F}$ corresponds to
a global optimisation problem over polynomially many parameters in an exponentially large space.
On the contrary, a least squares regression with $T$ training points, each with its own label, is asymptotically $\mathcal{O}(T^{3})$ in complexity.
However, in order to perform the regression, a training set of the random entanglement spectra of size $2^N$, needs to be created as described in Appendix \ref{app:spectracreation}.
The complexity of this is $\mathcal{O} ( {2^{3N}} )$.
Labelling the training points requires performing the minimisation in Ref~\onlinecite{Turner2017}
whose runtime scales as the time required to perform a local optimisation over the $\mathcal{O}(N)$ optimisation parameters multiplied with the number of basins visited.
After training, the prediction of $D_\mathcal{F}$ corresponds to
matrix multiplication, whose complexity is cubic in the matrix dimension.

\subsection{Abelian Topological States}\label{sec:AbelianTopoStates}

We now turn to
topological models at their renormalisation flow fixed point.
In Ref.~\onlinecite{KonMei18},
parafermionic chains (1D) and abelian string-nets ($2$ or higher D)
are studied
due to the specific structure of their ground state entanglement.
Before continuing, we reiterate these results,
as they are important for motivating the linear learning model we introduce.

\subsubsection{Optimal Free Spectra for Flat Spectra}

The entanglement spectra obtained from such states
comprise of only one eigenvalue with some degeneracy
dictated by the instance of the model.
In particular, for $\mathbb{Z}_N$ parafermions,
the entanglement spectrum from an equipartition of the chain in its topological phase
consists of an $N$-fold degenerate eigenvalue~\cite{Fendley:2012hw},
equal to $\frac{1}{N}$,
where we denote with overbar such flat spectra, $\bar P_N$.
For $\mathbb{Z}_N$ Abelian string-nets, the flat spectrum arising from bipartitioning the system
into a connected region and its complement has degeneracy $N^{|\partial|-1}$,
where $|\partial|$ is the length of the partition boundary~\cite{Alex}.
Thus determining the interaction distance reduces to determining the interaction distance for a parafermion case of the appropriate order.

It is conjectured that the free-fermion spectrum $P^\mathrm{f}_N$ closest to a flat spectrum $\bar P_N$ is constructed by
exactly reproducing as many of the highest probability eigenvalues of the flat spectrum as possible.
First, we pad $\bar P_N$ with zeros so that its size is equal to $2^{n+1}$,
where $n$ is the greatest integer such that $2^n \le N$,
so that the two spectra can be compared since in general their rank can be different.
Padding with zeros does not alter the entanglement and can be understood as introducing redundant unentangled degrees of freedom~\cite{Turner2017}.
The most that can be reproduced exactly are $2^n$, 
Then there exists one non-trivial fermion mode whose gap is fixed by normalisation.
In particular, in terms of Eq.\rf{eq:freespec},
for $n$ of the modes we set $s_i=0$ and one mode acquires gap $s_{n+1} = N^{-1}2^n - 1 / 2$.
Then, the optimal free fermion spectrum for an $N$-fold degenerate spectrum
\begin{equation}
  \label{eq:optimalguess}
  P^\mathrm{f}_N = \left(N^{-1},\dots,N^{-1},p,\dots,p \right),
\end{equation}
where there are $2^n$ entries for each value $N^{-1}$,
with $p = 2^{-n} - N^{-1}$ such that $\sum_k {P^\mathrm{f}_N}_k=1$.

Then evaluating $D_\mathcal{F}$ for such a choice of free spectra is straightforward.
There are two contributions.
The first is from the entanglement levels with index $2^n + 1 \le k \le N$ for which the probability difference is between $N^{-1}$ and $p$.
The second is from levels with index $N + 1 \le k \le 2^{n+1}$ for which the probability difference is between $0$ and $p$.
Thus we obtain
\begin{equation}
D_\mathcal{F}(\bar P_N) \le 3 - \frac{N}{2^n} - \frac{2^{n+1}}{N}.
\label{eq:upperbound}
\end{equation}
This result constitutes an upper bound, since the construction of the free spectrum is a conjecture.
However, numerical evidence partly shown in Fig.\ref{fig:hills} and fully supported in Ref.~\onlinecite{KonMei18}
supports that the equality holds
and thus we will assume this is the case.

The interpretation of the minimum and maximum value of $D_\mathcal{F}(\bar P_N)$ is as follows.
In the trivial case where $N=2^n$ for some $n\in\mathbb{N}$,
then this flat spectrum can be reproduced by $n$-many gapless fermion modes,
$\bar P_{2^n}=P^\mathrm{f}_{2^n}$,
and it is free with $D_\mathcal{F}=0$.
This result has important implications in studying free-fermion parent Hamiltonians of these topological models~\cite{KonMei18}.
By analytical continuation of $\mathbb{N}$ we set $N \rightarrow \alpha 2^n$ with $\alpha\in[1,2]$ such that we densely cover the interval between consecutive powers of two between which $N$ lies. 
Then, maximising Eq.\rf{eq:upperbound} we find $D_\mathcal{F}^\text{max}=3 - 2\sqrt{2}$
with
$\mathrm{argmax} \left( D_\mathcal{F}(\bar P{(\alpha)}) \right) = \sqrt{2}$.
The fact that the maximum occurs at an irrational value means that no flat spectrum can instantiate
the maximal value of $D_\mathcal{F}$. However, it can be approximated arbitrarily by the appropriate choice of $N$.
By the exhaustive numerical maximisation $\max_{P} D_\mathcal{F}(P)$ for random spectra
of size up to $2^8$ we have not found states with interaction distance larger than $D_\mathcal{F}^\text{max}$. Hence, this appears to be the maximum possible value of the interaction distance \emph{for any spectrum}.

\subsubsection{Supervised Linear AutoEncoder}

We now turn to supervised linear method inspired by
autoencoders for learning optimal free spectra $P^\mathrm{f}_N$
for flat spectra $\bar P_N$.
Again, we make a linear approximation to the expand map as in Eq.\rf{eq:linearexpand}.
First, we define a linear regression problem which
qualifies as a supervised learning protocol.
We fix an orthonormal basis
$e_n$, with $n=1,\dots,n_\mathrm{max}$,
viewed as carrying labels of \emph{free} flat spectra $\bar P_{2^n}$
all of which are
padded with zeros so that all of them are of size $2^{n_\mathrm{max}}$ and sorted in descending order.
Thus, we can consider a
design matrix $F=(\bar P_{2^1},\bar P_{2^2},\dots,\bar P_{2^{n_\mathrm{max}}})$
for which $\left|\left| e_n - F \bar P_{2^n} \right|\right|_2 =0$.
This matrix is full column rank and its
Moore-Penrose pseudoinverse ${F}^g$
is the solution of this linear regression problem.
This problem can be viewed as one of Independent Component Analysis~\cite{ICA},
where the uncorrelated sources that need to be discerned
correspond to the single-particle levels and the mixing matrix corresponds to the expand map.

We then
define a supervised linear autoencoder (SLAE)
as a modification of Eq.\rf{eq:autoencoder},
where we fix a basis $e_n$ for the latent layer
$\mathcal{C}(X)$.
Furthermore,
the Coding and Decoding processes correspond to
solutions of a linear regression problem $F X = e_n$,
whose design matrix
effectively implements $\mathcal{C}=F$ and $\mathcal{D}=F^g$.
In this respect, the corresponding interaction distance for an entanglement
spectrum $Y$ from a test set is defined as
\be\label{eq:DF_SLAE}
D^\mathrm{SLAE}_\mathcal{F}=D(F^g F Y,Y).
\ee

\begin{figure}[t]
\centering

\includegraphics[width =.92\columnwidth]{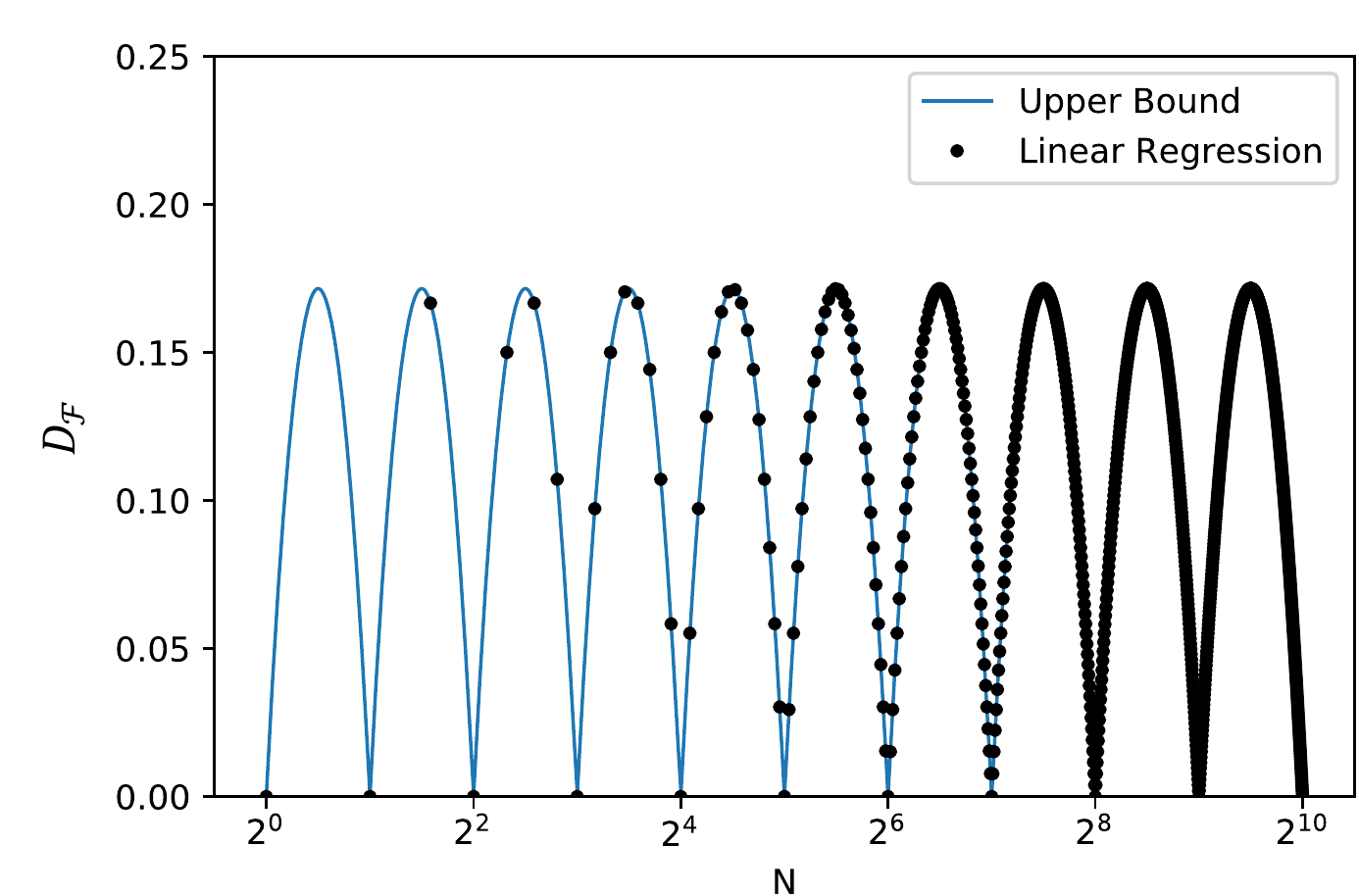}

\caption{Interaction distance for $N$-rank flat spectra.
The dots are the prediction $D^{SLAE}_\mathcal{F}(\bar P_N)$ of the Supervised Linear AutoEncoder. It agrees with both the numerical results for the interaction distance
in Ref.~\onlinecite{KonMei18}
and the upper bound of its analytical form of Eq.\rf{eq:upperbound}.
}
\label{fig:hills}
\end{figure}

For the particular problem of computing $D_\mathcal{F}(\bar P_{N})$,
the matrix
$F$ contains free flat spectra $\bar P_{2^n}$
and is identified with
the Moore-Penrose inverse
of the linear version of the expand map, $\slashed{\mathcal{E}}^g$.
Note that $F$ here is full column rank.
Then the SLAE
predicts the interaction distance
for all $N$-rank flat spectra $\bar P_N$
as $D^{SLAE}_\mathcal{F} (\bar P_N)= D(  \mathcal{D}\mathcal{C} \bar P_N , \bar P_N )$,
shown in Fig.~\ref{fig:hills}.
The fact that $D^{SLAE}_\mathcal{F}=D_\mathcal{F}$
means the guess for the optimal free spectrum corresponding
to a flat spectrum of Eq.\rf{eq:optimalguess}
can be viewed as a linear combination of free flat spectra.
In Appendix \ref{sec:SLAEFuzzyFlat} we show that the prediction is robust for almost flat spectra.
Since the training set for this specific case (with $n_{\mathrm{max}} = 10$) is limited to 10 basis states, the calculation of the design matrix and its pseudoinverse is almost instantaneous. A massive speedup from the general $D_{\mathcal{F}}$ algorithm which takes $\sim$3 hours for $n_{\mathrm{max}} = 10$ and $100$ basins visited for each $\bar P (N)$.


\subsection{Quantum Ising Chain} \label{sec:IsingSLAE}

Finally, we
relate the extensive results of Ref.~\onlinecite{Turner2017} on the interaction distance
for ground states of the antiferromagnetic (AFM) quantum Ising chain
with our supervised linear autoencoder method.
The Hamitlonian $H(h_z,h_x)$, explicitly written in Eq.\rf{eq:IsingHam} of Appendix \ref{sec:IsingApp}, has a parameter line $h_x=0$ on which the model $H(h_z,0)$ maps to a free-fermions with a quantum critical point at $h_z=1$.

By equipartitioning the chain, we obtain the entanglement spectrum from the ground state.
The ground state is obtained by exact diagonalisation. One can employ matrix-product states
to access larger chain lengths.
We denote the entanglement probability spectra obtained by equipartitioning the chain in its ground state as $P(h_z,h_x)$, with corresponding entanglement energies $E(h_z,h_x)=-\log{P(h_z,h_x)}$.
On the free line we have $D_\mathcal{F}(P(h_z,0))=0$, $\forall h_z$.
In Ref.\onlinecite{Turner2017} it is demonstrated by a detailed scaling analysis
that in the thermodynamic limit the model is almost free everywhere in its phase diagram.
For a finite system size we have finite $D_\mathcal{F}$ on the critical line.
Finally, we map each $P(h_z,h_x)$ on the phase diagram to the free line
by minimising its distance from that line, $\min_{h_z^\mathrm{f}}D(P(h^\mathrm{f}_z,0),P(h_z,h_x))$. The field value $h^\mathrm{f}_z$ for which the minimum occurs characterises isofree spectra.

\begin{figure}[t]
\centering 
\includegraphics[width =.92\columnwidth]{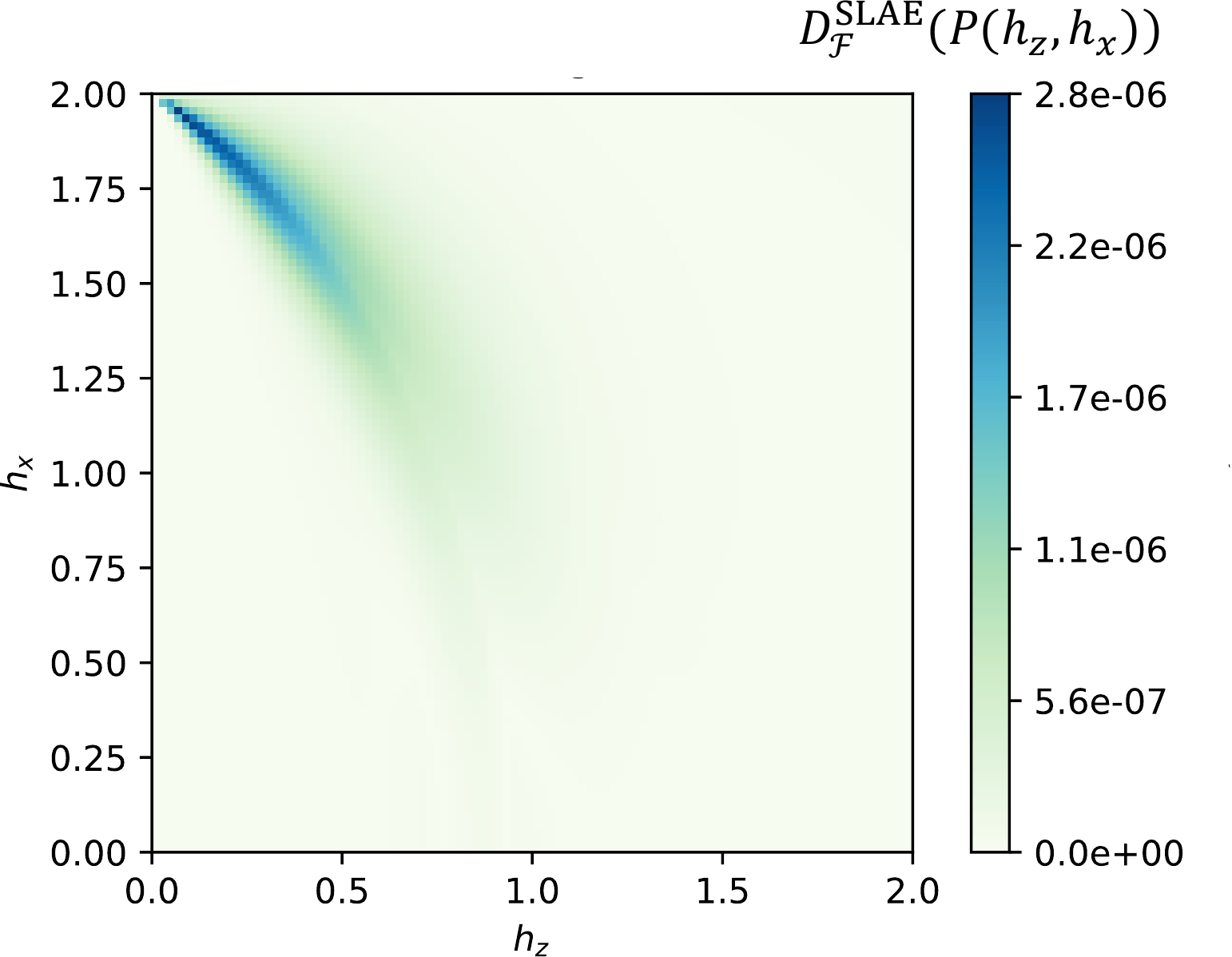}
\includegraphics[width =.92\columnwidth]{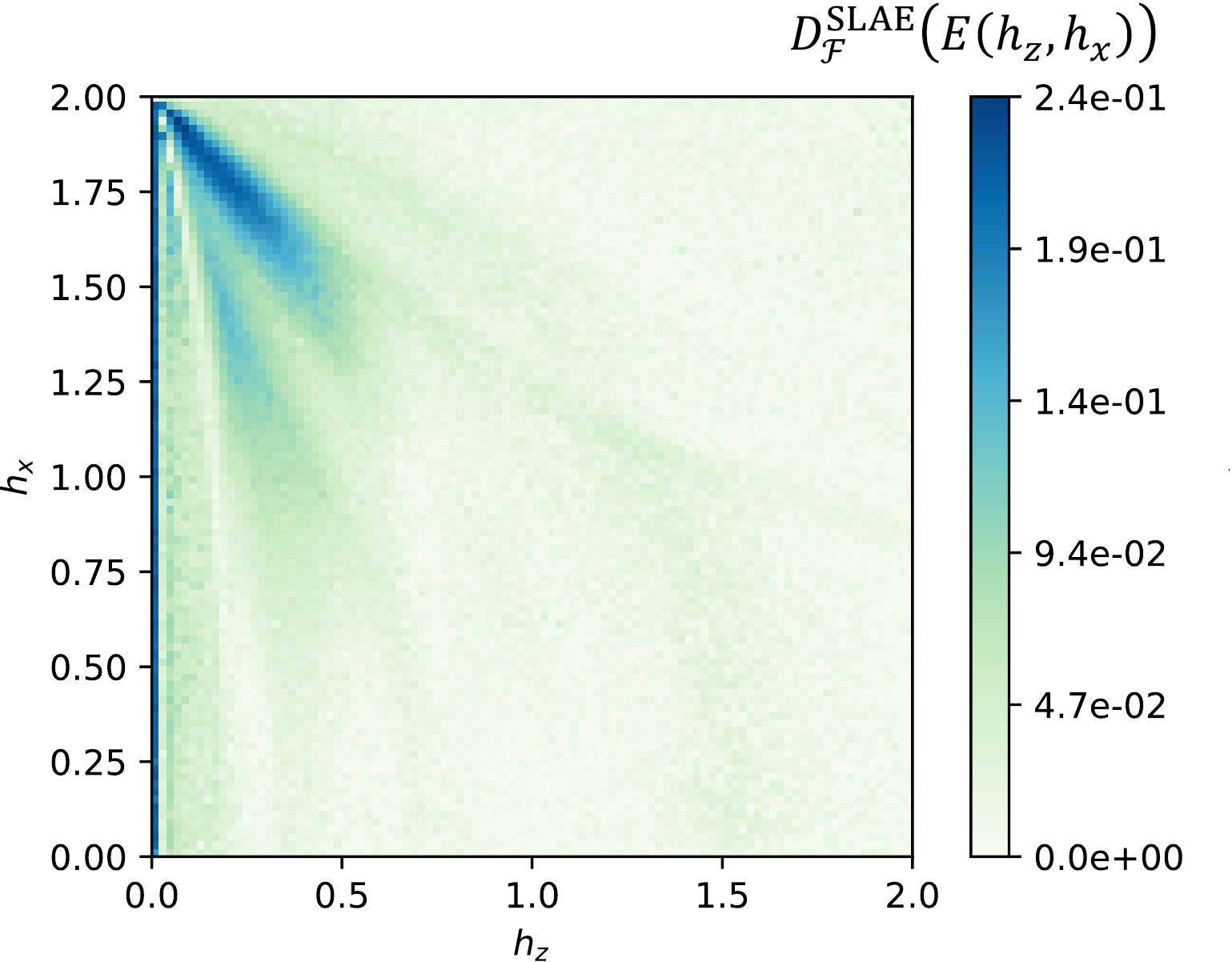}

\caption{Estimated interaction distance $D_\mathcal{F}^\mathrm{SLAE}$
with the supervised linear autoencoder trained in probability space $P(h_z,h_x)$ (left) or entanglement energy space $E(h_z,h_x)$ (right).
The colourmap is in correspondence with that of Fig.3 in Ref.~\onlinecite{KonMei18}.
}
\label{fig:SLAE_Ising}
\end{figure}

We implement the SLAE,
as it is defined in Eq.\rf{eq:DF_SLAE},
in order to quantify the distance between any $P(h_z,h_x)$ spectrum
from the closest spectrum $P(h^\mathrm{f}_z,0)$ on the free line.
The size $n_\mathrm{max}$ of the basis $e_n$ for the latent layer is in this case a parameter we must determine. This is done by analysing how well the entire free line is reproduced by each size of basis by minimising the absolute error of the free line over $n_\mathrm{max}$ and adding a penalty for increased $n_\mathrm{max}$ as
$\min_{n_\mathrm{max}} \sum_{n=1}^{n_\mathrm{\max}} D(\mathcal{D}\mathcal{C}P(h_z(n),0), P(h_z(n),0))  + \alpha n $. Where $\alpha$ is a constant of the order of the absolute error.
The basis elements correspond to labels for free spectra
$P(h_z (n),0)$
that are chosen
as representatives for the free line for $n$-many values of $h_z$.
These free spectra then enter the columns of the design matrix
of the linear regression problem of the SLAE.
Then, the predicted distance from the free line is $D^{SLAE} (P(h_z,h_x))= D(  \mathcal{D}\mathcal{C} P(h_z,h_x) , P(h_z,h_x) )$.
In Fig.~\ref{fig:SLAE_Ising} we demonstrate that the results correspond
to the equivalent results of Ref.~\onlinecite{Turner2017}.


\section{Autoencoder}

Finally, we
present an alternative method of producing results equivalent to those
obtained with SLAE for the flat spectra and the AFM Quantum Ising Chain
in Sections \ref{sec:AbelianTopoStates} and \ref{sec:IsingSLAE}.
The AutoEncoder (AE) in this case is semi-supervised and is trained on
free spectra relevant to the particular problem at hand.
Following the discussion in Section \ref{sec:randomstates} on the different behaviour
of a linear method on the probability and energy spaces,
we are motivated by the fact that a promising proposal for a neural network that would outperfrom the linear approaches we take here should be designed to overcome these limitations. It should both capture the geometry of $\mathcal{F}$ and have an appropriate surrogate for the trace distance cost function. It should also naturally operate on multisets such that it is not confused by reorderings in the dataset.

Our autoencoder is built in TensorFlow and consists of an input/output of $\mathcal{I} = 2^N$ neurons, with three hidden layers, $h_{1},h_{2},h_{3}$.
The latent vector, $h_2$, is chosen to have size $\mathcal{L} = N$, such that a free system can be fully described by these neurons.
Various activation functions were tested on all layers, however the best results were found when only a softmax~\cite{bishop2006pattern} activation function was applied to the output layer. This has the benefit of enforcing normalisation of the output.

\begin{figure}[t]
\centering

   \includegraphics[width=.88\columnwidth]{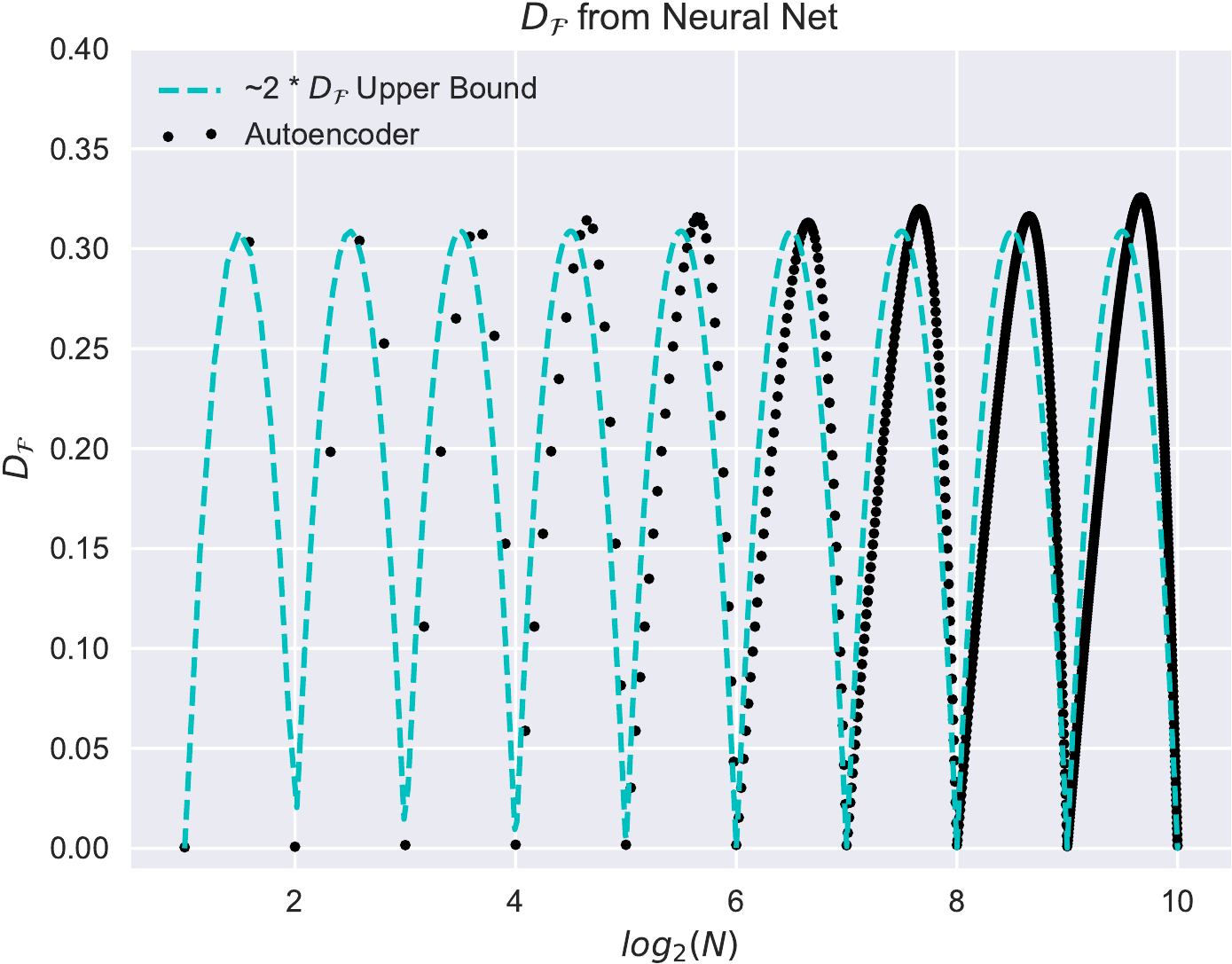}         
      \includegraphics[width=.9\columnwidth]{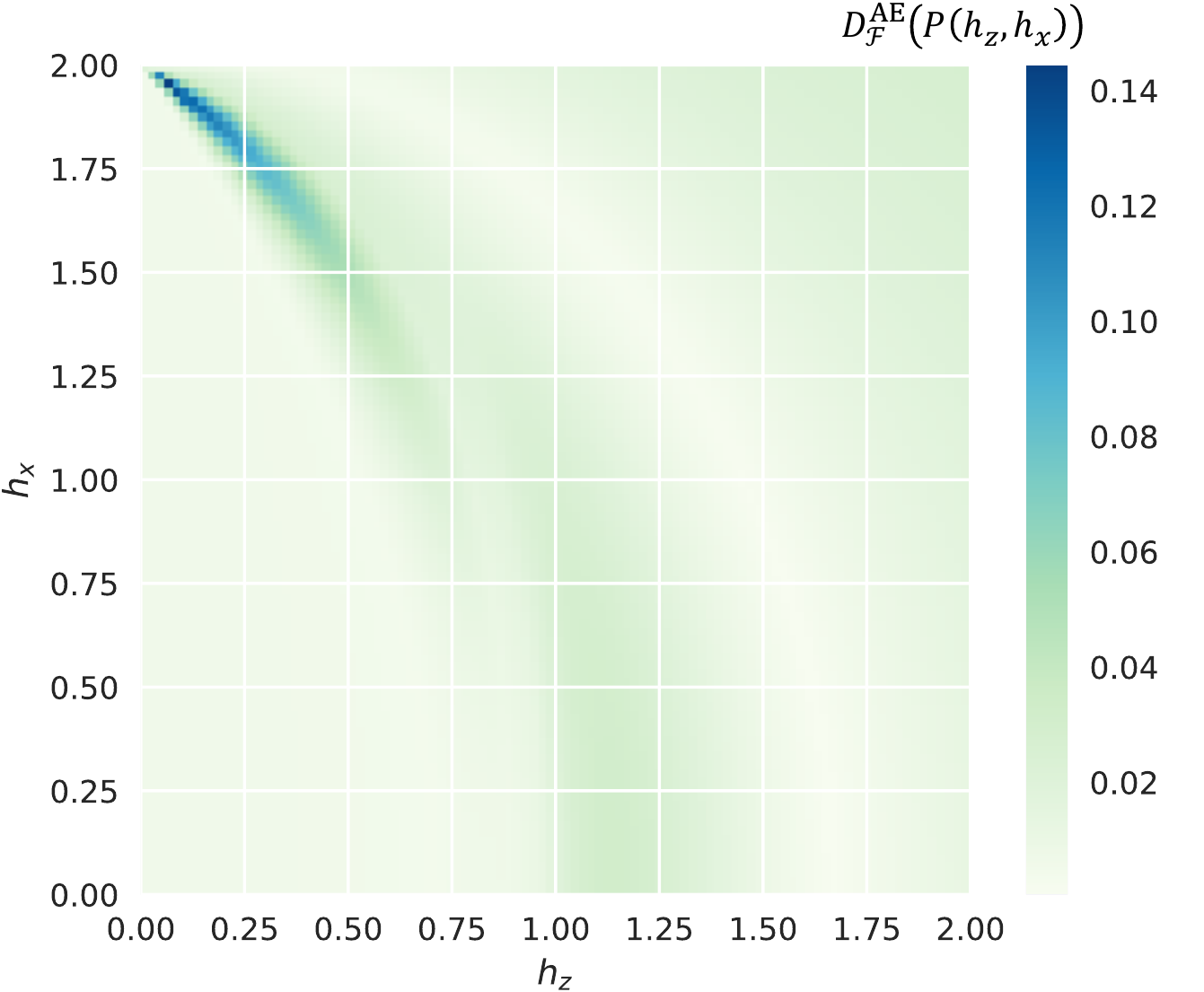}         
  \caption{ (Top) Autoencoder estimation $D^\mathrm{AE}_\mathcal{F}(\bar P_N)$ trained on almost flat spectra $P^\Delta_N$ (see Appendix \ref{sec:SLAEFuzzyFlat}). We obtain a log-periodic curve which is viewed as a surrogate for $D_\mathcal{F}(\bar P_N)$. Here $\mathcal{I}=1024$ and $\mathcal{L}=10$.
(Bottom) $D^\mathrm{AE}_\mathcal{F}(P(h_z,h_x))$ for AFM Ising, trained on the free spectra $P^\Delta(h_z,0t)$ with $\Delta=0.1$. Here, chain length is $L=10$, and we sampled $10^4$ values of $h_z$. Here $\mathcal{I}=32$ and $\mathcal{L}=5$.}
  \label{fig:AE_Hills_AFM}
\end{figure}

In the case of flat spectra, obtained from Abelian topological states,
the AE is trained on flat and almost flat spectra as they are defined in Appendix \ref{sec:SLAEFuzzyFlat}, that is flat spectra with disordered eigenvalues.
This is done to increase the training set's size, and for low disorder amplitudes it is expected
to not affect the performance of the AE.
We observe
in Fig.\ref{fig:AE_Hills_AFM} (top)
that we reproduce
a log-periodic function for the interaction distance.
The shape of the curve need not be that of Eq. \rf{eq:upperbound}.

For the quantum Ising Chain, entanglement
spectra $P^\Delta(h_z,0)$ are sampled from the free line $h_x=0$
and comprise the training set.
The sample size is increased by introducing disorder of amplitude $\Delta$ in the couplings
of the chain as described in Appendix \ref{sec:IsingApp}.
For this problem, the network was seen to overfit to the free line and therefore identifies the majority of the phase diagram as strongly interacting.
To battle this, dropout regularization~\cite{Dropout} was applied to $h_3$, with a dropout probability of $P_\mathrm{D} = 0.5$.
This significantly improved the fitting of our AE to the majority of the phase diagram, which is shown in Fig. \ref{fig:AE_Hills_AFM} (bottom).

The goal of the training is for the AE to effectively learn the map $\mathcal{E}\mathcal{E}^g$
as described in Sec. \ref{sec:AEPerspective},
from the dataset of free states we provide
for the specific case studies at hand.
Note that in contrast to the SLAE, we do not fix the basis of the vector space
corresponding to the latent layer of the AE;
the AE learns an input-output relation by trying to reproduce the input it is given
as the output of a compression-decompression process implemented
by a deep neural network. Limiting the number of neurons in the hidden layer of the neural network and forcing a dimensionality reduction forces the network to learn \emph{features} of the training data that are robust to variation.

\section{Conclusions and Outlook}

We have applied 
machine learning techniques to compute the interaction distance.
By choosing to do so, we exchange generality with performance.
In particular, the algorithm presented in Ref.~\onlinecite{Turner2017}
for $D_\mathcal{F}$ can be applied to any spectrum.
Our linear methods estimate $D_\mathcal{F}$
considerably faster but with the loss of generality due to the need of retraining for each model.
Regarding the autoencoder, it can in principle be trained to perform
a task analogue of estimating $D_\mathcal{F}$. The estimation step is indeed faster than our original brute force optimisation. However,
the training set required is of high complexity.
We leave it as future work to explore sampling gMPS~\cite{gMPS}, gMERA~\cite{Fishman,MERAwavelets}, gRandomTensorNets~\cite{EvenblyWhite}, and eigenstates of random free-fermion Hamiltonians on random graphs, in order to create the dataset.

As future work,
we can formulate the problem of estimating the interaction distance in terms of
the correlation matrix of the state, that is its two-point correlations~\cite{Gertis:2016wj}.
In this case, the training dataset consists of correlation matrices obtained for free fermion systems.
Again, even if the generation of a free-fermion correlation matrix is efficient ($\mathcal{O}(N^3)$), the free fermion systems to which they correspond need to be good
representatives of $\mathcal{F}$.
The approach in this case is different, however.
Correlation matrices can be used as input to pattern recognition algorithms acting on images,
and we expect that for an appropriate dataset, the free-fermion structure is learned.
In any case, a significant improvement would be to have a method
which is insensitive to the sizes of the input spectra or correlation matrices.
In the case of entanglement spectra we can always pad with zeros to reach any dimension
required.

Finally, we briefly comment on the distinction between
classical and quantum random spectra, shown as distinct clusters
in Fig.\ref{fig:PCAspectra}.
Actually, the r-statistics of entanglement spectra comming from the former and the latter procedures obey the Poisson and Wigner Dyson GUE distributions, respectively\cite{Chamon}.
GUE is conjectured to signify universality of the quantum process that generated that state. Poisson is conjectured to signify non-universality.
For the purposes of this work, we call the former spectra classical
since the conjectures about the entanglement r-statistics are results
of studies on Clifford circuits which are classically simulable and universal circuits.
However,
there are examples of non-universal quantum processes that generate
states whose probability distributions are hard to sample from classically.
We leave it for future work to investigate whether states that are output of
such non-universal but classically hard to simulate processes
manifest as a cluster between the two aforementioned clusters.
One example would be boson sampling, where one would use the best known classical algorithm for computing permanents~\cite{Neville2017}. The entanglement r-statistics
are also an open question; that is whether another distribution Poisson or GUE.\\

{\bf Acknowledgements :}
We thank Zlatko Papic for helpful discussions.
C.J.T. acknowledges financial support by the EPSRC grant EP/M50807X/1.
K.M. acknowledges financial support by the EPSRC Doctoral Prize EPSRC1001.
Preprint of an article submitted for consideration in  the International Journal of Quantum Information \textcopyright (2018) copyright World Scientific Publishing Company (http://www.worldscientific.com/worldscinet/ijqi).

\pagebreak
\appendix

\begin{center}
{\bf APPENDIX }
\end{center}

Here we describe the procedures of obtaining the spectra we study in the main text.
Our constructions are either direct construction of the spectrum, or sampling ground states of specific Hamiltonians.
Furthermore, we show evidence that the SLAE is stable under
input of almost flat spectra.

\section{Constrution of Entanglement Spectra}
\label{app:spectracreation}

We describe in detail how we construct 
random free, generic random, abelian topological, and free Ising
entanglement spectra.
The PCA of these spectra is shown in Fig.\ref{fig:PCAspectra} in the main text.

\subsection{Random Free Spectra}

A random free entanglement spectrum are sampled as follows.
First sample single-body $s_i$ or $\epsilon_i$ form a probability distribution of our choice
and obtaining a many-body spectrum of the form of Eq.\rf{eq:freespec} via the expand map.

\subsection{Random Spectra}

We call classical spectra random probability distributions
constructed by sampling random real numbers $P_k\in [0,1]$ and normalising them as $P_k \rightarrow P_k / {\sum_k P_k}$.
On the other hand, we construct quantum random spectra as follows.
A random matrix of complex elements $M_{ij} \in \mathbb{C}$ is created, with
$i=1,\dots,2^n$ and $j=1,\dots,2^m$. It corresponds to
the entanglement matrix of a random $(n+m)$-qubit state with amplitudes $\psi_k$
defined when one attempts the Schmidt decomposition of said state,
$\ket{\psi} = \sum_{ij} M_{ij} \ket{i}\ket{j}$.
Then the entanglement spectrum is obtained by $\xi=\mathrm{svd}(M)$.

In particular, we sample random complex numbers for the elements of $M$
as $M_{ij}=\alpha_{ij} e^{i \beta_{ij} 2\pi}$,
where $\alpha_{ij}\in P$ random numbers from some probability distribution of our choice
and $\beta_{ij}\in[0,1]$ uniform random numbers.
We restrict to real states by setting $\beta_{ij}=0$.
Then we normalise $M_{ij}\rightarrow M_{ij} / \tr{M^\dagger M}$.

\begin{figure}[htb]
\centering
  \includegraphics[width=.9\columnwidth]{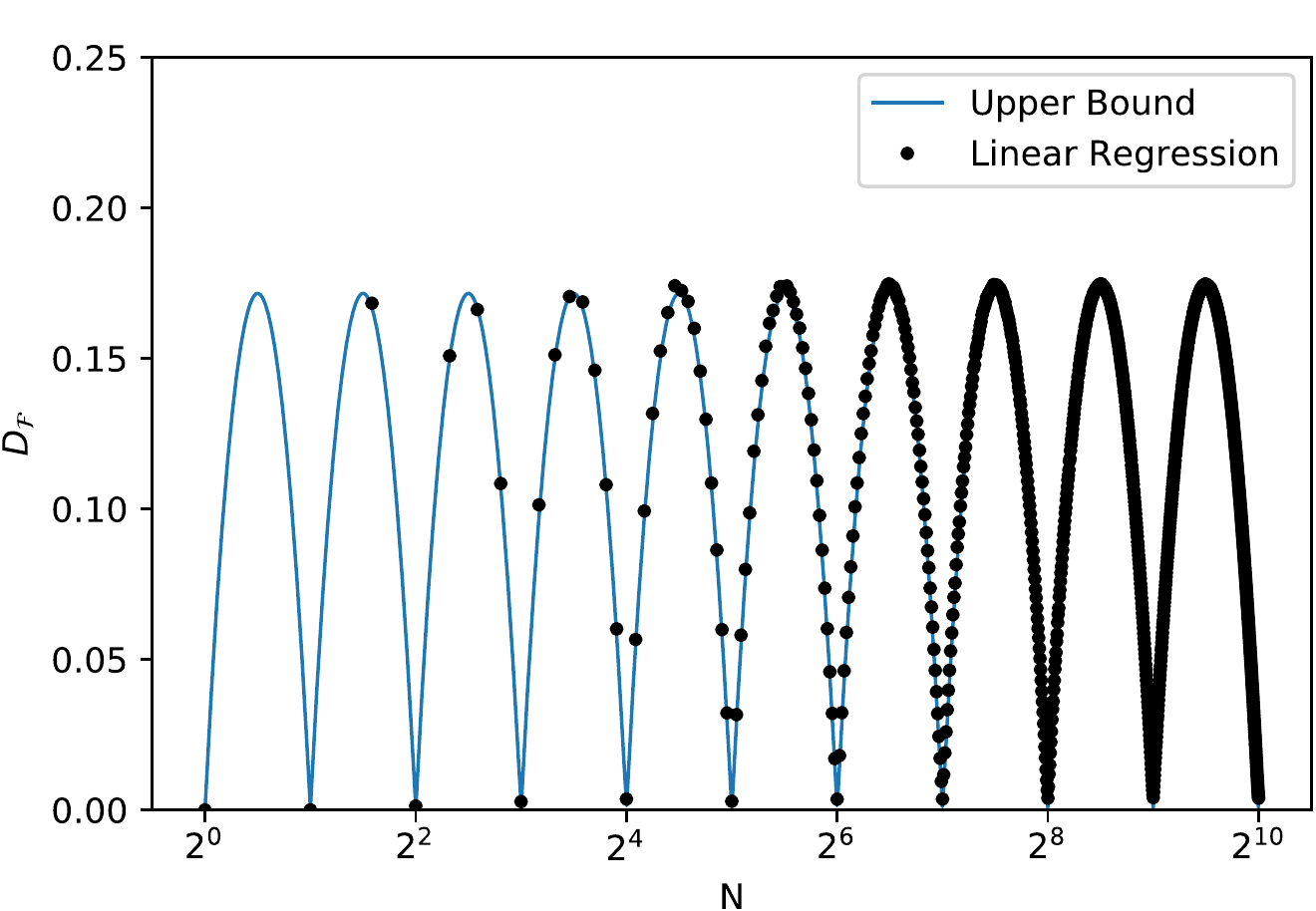}
  \caption{ $D^{SLAE}_\mathcal{F}(\bar P(N))$ for almost flat spectra with disorder of amplitude $\Delta \approx 0.01$ }
  \label{fig:SLAEFuzzyFlat}
\end{figure}

\subsection{Parafermion Spectra}
The 1D Ising model can be mapped to the Majorana chain by means of a Jordan-Wigner transformation~\cite{KitaevChain}.
Similarly, $\mathbb{Z}_{N>2}$ generalisations of the Ising model known as the clock Potts model can be expressed in terms of parafermions~\cite{FradkinKadanoff,Fendley}.

The chains are described by the Hamiltonian
\begin{equation}
  \label{eqn:parafermions}
  H_{\mathbb{Z}_N} = - e^{i\phi}\sum_j \alpha^\dagger_{2j} \alpha_{2j+1} - f e^{i\theta}\sum_j \alpha^\dagger_{2j-1} \alpha_{2j} + \text{h.c.},
\end{equation}
where  $f$ is real and $\phi$ and $\theta$ are the chiral phases of the model. The parafermion operators satisfy the generalised commutation relations $\alpha_j \alpha_k = \omega \alpha_k \alpha_j $ for $k>j$, where $\omega = e^{i 2\pi/N}$ and $(\alpha_j)^N=1$.
Here we focus on the gapped regime away from the critical points or critical phases~\cite{Elitzur1979}. 

Fixed point: $f=0$ with $\phi=\theta=0$ and we place the bipartition between regions $A$ and $B$ at a $(2j,2j+1)$-link. The obtained spectrum is an $N$-fold degenerate eigenvalue understood as the number of parafermionic parity admissible to each partition and compatible with the global fixed parity of the chain.
Our analysis is non-trivial as the fixed-point results are robust off-the-fixed point ($f>0$), as well as for finite chirality $\phi,\theta\neq 0$~\cite{KonMei18}.

\subsection{Ising Spectra}\label{sec:IsingApp}

We obtain entanglement spectra from equipartitioning ground states of the quantum Ising chain with periodic boundary conditions whose Hamiltonian is
\be \label{eq:IsingHam}
H (h_z,h_x)=-\sum_{j=1}^L  J X_j X_{j+1} +h_z Z_j + h_x X_j ,
\ee
where $J>0$ and $j$ runs over sites on which $\frac{1}{2}$-spins are defined
and $X_j,Z_j$ are Pauli matrices that act on those spins.
The boundary condition $L+1=1$ ensures that the ground state is unique.

In this work we set $J=1$.
To sample free Ising spectra we set $h_x=0$.
On this free line of the phase diagram the chain can be mapped to free fermions via the Jordan-Wigner transformation.
Note that this free-fermion model is the well studied Kitaev chain which corresponds to the $\mathbb{Z}_2$ parafermion chain.
To further increase the sample size we can also introduce disorder in
the couplings, $J \rightarrow J+r$ and $h_{x,z}\rightarrow h_{x,z}+r$, where $r$ is a Gaussian random number $\mathcal{N}(0,\Delta)$. We call $\Delta$ the disorder amplitude.
To sample from off-the-free-line we simply turn on the longitudinal field $h_x>0$. Disorder can be introduced in the same way to increase the sample size.

\section{Supervised Linear Autoencoder on Almost Flat Spectra} \label{sec:SLAEFuzzyFlat}

The predicted interaction distance
of the supervised linear autoencoder is robust under perturbations of the input flat spectrum.
Such perturbed spectra we refer to as almost flat, denoted $\bar P^\Delta_N$.
Each non-zero eigenvalue is of the form
$\Delta \frac{1}{N} $ and normalisation is ensured by dividing by their sum.
For almost flat input spectra the prediction $D^{SLAE}_\mathcal{F}$ is shown in Fig.\ref{fig:SLAEFuzzyFlat}.

\bibliographystyle{ws-ijqi}

\end{document}